\documentclass[useAMS,usenatbib]{mn2e}
\usepackage{graphicx}
\usepackage{amsmath}
\usepackage{tabularx}

\usepackage{times}
\usepackage{amssymb,amsmath}

\def\lsim{ \lower .75ex\hbox{$\sim$} \llap{\raise .27ex \hbox{$<$}} }
\def\gsim{ \lower .75ex \hbox{$\sim$} \llap{\raise .27ex \hbox{$>$}} }

\newcommand{\bi}{\begin{itemize}}
\newcommand{\ei}{\end{itemize}}

\title[Time-dependent X-ray polarimetry of BL Lac jets] 
{Probing shock acceleration in BL Lac jets through X-ray polarimetry: the time-dependent view} 

\author[Tavecchio et al.]
{F. Tavecchio$^1$\thanks{E--mail: fabrizio.tavecchio@inaf.it}, M. Landoni$^1$, L. Sironi$^2$  and P. Coppi$^3$
\\
$^1$INAF -- Osservatorio Astronomico di Brera, via E. Bianchi 46, I--23807 Merate, Italy\\
$^2$Department of Astronomy, Columbia University, 550 W 120th St, New York, NY 10027, USA\\
$^3$Yale Center for Astronomy and Astrophysics, Yale University, New Haven, CT 06520-8121, USA
}

\begin{document}



\maketitle

\begin{abstract} 
Polarimetric measurements, especially if extended at high energy, are expected to provide important insights into the mechanisms underlying the acceleration of relativistic particles in jets. In a previous work we have shown that the polarization of the synchrotron X-ray emission produced by highly energetic electrons accelerated by a mildly relativistic shock carries essential imprints of the geometry and the structure of the magnetic fields in the downstream region. Here we present the extension of our analysis to the non-stationary case, especially suitable to model the highly variable emission of high-energy emitting BL Lacs. We anticipate a large ($\Pi \approx 40\%$), almost time-independent degree of polarization in the hard/medium X-ray band, a prediction soon testable with the upcoming mission {\it IXPE}. The situation in other bands, in particular in the optical, is more complex. A monotonic decrease  of the optical degree of polarization is observed during the development of a flare. At later stages $\Pi$ reaches zero and then it starts to increase, recovering large values at late times. The instant at which $\Pi=0$ is marked by a rotation of the polarization angle by $90$ degrees. However, at optical frequencies it is likely that more than one region contributes to the observed emission, potentially making it difficult to detect the predicted behavior.
\end{abstract}


\section{Introduction}

Electromagnetic emission from extragalactic jets (e.g. Romero et al. 2017, Blandford et al. 2019), extending over the entire spectrum, from the radio band up to gamma rays, is produced by non-thermal populations of relativistic particles carried by the outflowing plasma. Despite the intense efforts devoted to the topic, the nature of the basic mechanism responsible for the acceleration of these particles is still unclear and debated. While diffusive shock acceleration (DSA, e.g. Blandford \& Ostriker 1978) has been considered for a long time as the most natural candidate, recent studies have highlighted the important role potentially played by magnetic reconnection (e.g. Sironi et al. 2014, Guo et al. 2014, 2015, Werner et al. 2016), turbulence (e.g. Zhdankin et al. 2017, 2019; Comisso \& Sironi 2018) or the interplay between the two (e.g. Comisso \& Sironi 2019).

Relativistic jets are best studied in blazars, active galactic nuclei where the jet is closely pointed towards the Earth (e.g. Urry \& Padovani 1995). Under this favourable geometry the non-thermal emission of the jet is highly amplified by relativistic boosting effects and it can easily swamp the thermal contribution from the active nucleus. The emission from blazars is characterized by a spectral energy distribution (SED) with two broad bumps (e.g. Fossati et al. 1998, Ghisellini et al. 2017). The low energy component is produced through synchrotron radiation by relativistic electrons (or pairs), while the high-energy peak is widely interpreted in terms of inverse Compton scattering by the same electron population (e.g. Maraschi et al. 1992, Sikora et al. 1994), although hadronic or lepto-hadronic models have also been  proposed (e.g. Boettcher et al. 2013, Cerruti et al. 2015). 
From the point of view of the particle acceleration, particularly interesting are the so-called highly peaked BL Lacs (HBL). The SED of these blazars has maxima in the X-ray and at TeV energies, pointing to the presence of particles  pushed to extremely high energies (e.g. Tavecchio et al. 2010). 

For a long time, multiband polarimetric measurements have been considered a powerful investigation tool of jet structure and dynamics (e.g. Angel \& Stockman 1980, Blandford et al. 2019). In the last decade, the regular multiband monitoring of blazars, sparked by the advent of {\it Fermi}-LAT, led to the identification of possible regular patterns involving polarimetric properties. In particular, the evidence for systematic and large variations of the polarization angle, potentially associated to powerful gamma-ray flares (e.g., Blinov et al. 2015,2018), has been interpreted in terms of an emission region moving along a helical path in a jet dominated by a toroidal field (e.g. Marscher et al. 2008, 2010, Larionov et al. 2013) or as caused by a jet bending at parsec scales (Abdo et al. 2010, Nalewajko 2010). However, similar features can also be explained by scenarios where the polarization behavior is related to turbulence in the flow (usually described in terms of stochastic models, e.g., Kiehlmann et al. 2016, 2017), possibly generated downstream of a standing shock (Marscher 2014, 2015). In this last framework the observed emission does not carry any direct information on the structure of the magnetic field in the jet, since its properties are mainly shaped by the turbulent nature of the flow.

From the theoretical perspective, two main routes are generally considered to convert part of the outflow energy flux to the population of energetic particles. For outflows characterized by large magnetization the most efficient and likely dissipation process is the {\it direct} conversion of magnetic energy into particle energy through magnetic reconnection. In fact, intense dissipation of magnetic energy through reconnection is expected to accompany the global reorganization of the fields triggered by kink instabilities (e.g. Begelman 1998, Giannios \& Spruit 2006, Barniol Duran et al. 2017). On the other hand, if dissipation occurs after the conversion of a sizeable fraction of the original magnetic energy to the bulk kinetic energy of the outflowing plasma, the low magnetization allows the formation of shocks and therefore a second possibility, i.e. diffusive shock acceleration. From the observational point of view, it is rather difficult to make a distinction between these two alternatives, since similar electron energy distributions (i.e. power laws) are expected in both cases. Fast variability could provide some clues, although, lacking a precise characterization of the phenomenology and the frequencies of such events, it is not clear whether flares with extremely short timescale (likely favouring magnetic reconnection scenarios, e.g. Christie et al. 2019) should be considered the normality or, rather, exceptional events. 

As discussed in Tavecchio et al. (2018, hereafter T18), acceleration through DSA is expected to imprint specific polarimetric signatures in the synchrotron emission of accelerated particles, especially in the case of HBLs. Particle-in-cell (PIC) simulations of trans-relativistic shocks have indeed demonstrated that the acceleration process proceeds through the formation of self-generated magnetic fields close to the shock front (T18, Vanthieghem et al. 2020, Crumley et al. 2019). In turn, the predominantly orthogonal self-generated fields result in a high degree of polarization for the X-ray emission, produced by high-energy electrons cooling very close to the shock front. The prediction of a large polarization of the X-ray emission is going to be soon tested by the upcoming {\it IXPE} satellite (Weisskopf et al. 2016). However, in view of the observational test of this scenario, it is mandatory to explore a  situation more realistic than the stationary case discussed in T18. In fact, a time-independent set-up is suitable to model quiescent, low flux states. On the other hand, HBL typically display strong and persistent variability, especially in the X-ray band. At these energies the typical cooling time of the electrons is expected to be shorter than the observed variability timescale, making a time-dependent model essential.

With these motivations, in the present paper we intend to extend the model developed in T18 including a time-dependent study of the polarization properties of shocks. The paper is organized as follows: in Sect. 2 we present the model and the numerical implementation, in Sect. 3 we present the results for several cases of flares and in Sect. 4 we discuss the results. 

\section{A time-dependent model for polarization from shocks}

\begin{figure*}
 \centering
 \hspace*{-0.0truecm}
 \includegraphics[width=1.05\textwidth]{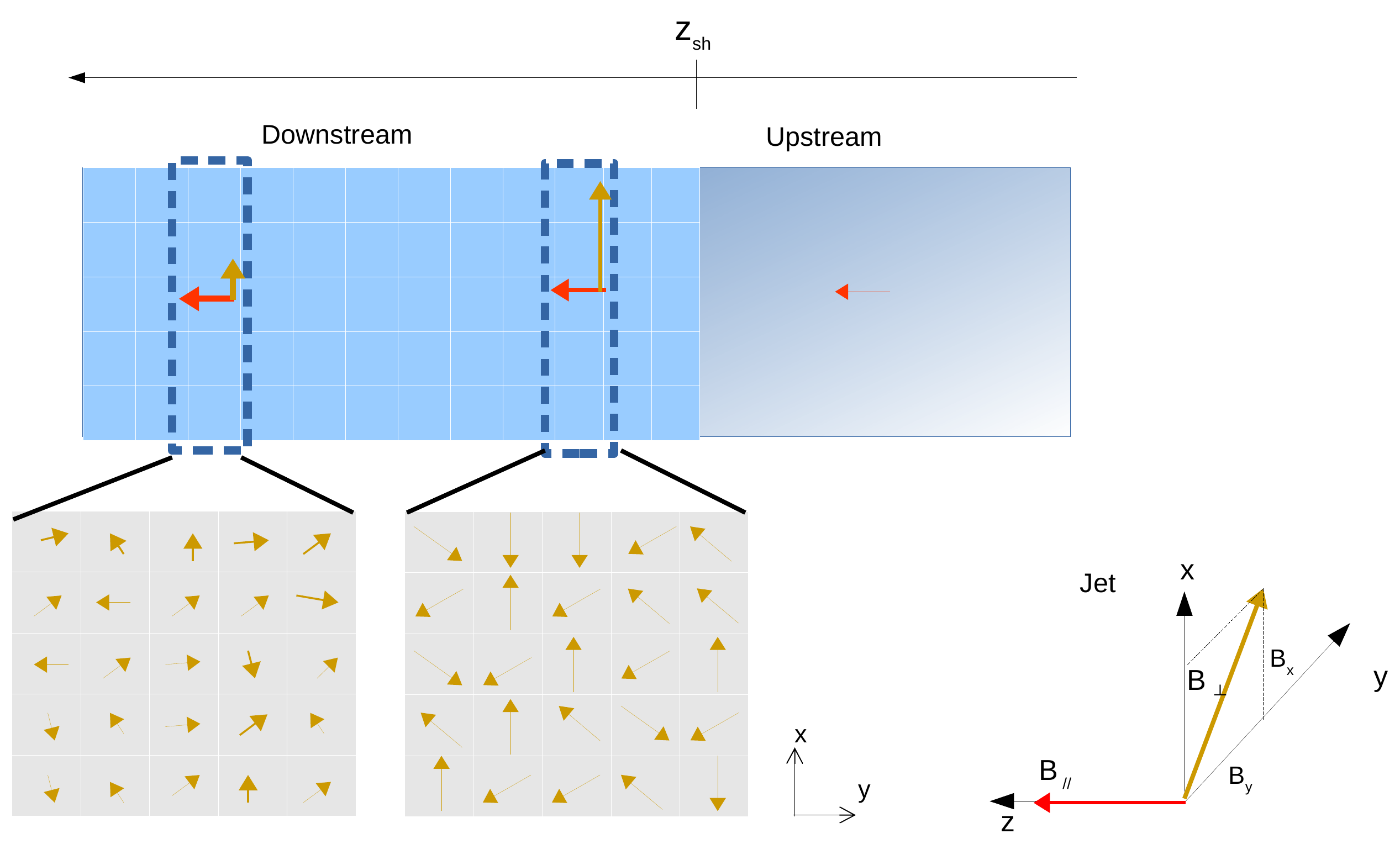}\\
 \caption{Sketch of the setup used. Plasma flows from the right to the left. The upstream flow (right), carrying a small poloidal magnetic field component (red arrow) reaches the shock (at $z=z_{\rm sh}$). At the shock front instabilities produce a predominantly orthogonal magnetic field (orange) whose structure is modelled by using cells characterized by different values of $B_{x}$ and $B_{y}$ (cross sections below). The line of sight in the downstream frame is aligned along the $y$ axis. Therefore $B_{x}$ corresponds to the projection of $\mathbf{B_{\perp}}$ on the plane of the sky. The self-generated field decays with distance from the shock. See text for more details.}
 \label{fig:cartoon}
\end{figure*}

In T18 we studied the properties of the polarization of  synchrotron radiation produced by relativistic electrons advected in the downstream region of a shock by means of a simple model inspired by PIC simulations. We assumed that the magnetic field displays two components, one quasi-parallel to the shock normal, carried by the upstream flow, and a perpendicular component, self-generated close to the shock front and rapidly fading downstream. The key feature of the above set-up is that electrons at the highest energies (contributing to the emission at the largest frequencies, in the X-ray band) rapidly cool and their emission occurs mainly in regions dominated by the orthogonal self-generated magnetic fields. On the other hand, particles at lower energies (emitting in the optical band) live for a time sufficient to reach a distance where the orthogonal component decreased below the value of the parallel component. In this configuration the X-ray emission, produced in a region with a well defined field orientation, is expected to be highly polarized, while the optical radiation, produced in regions where the magnetic field presents different orientations, is expected to display a lower degree of polarization.

In T18 we have studied the properties of the polarization in the stationary case (corresponding to equilibrium between injection and losses). Here we extend the previous stationary treatment to a fully time-dependent model, including light-travel effects. The time-dependence introduces new features with respect to the stationary scheme sketched above. In the next section we describe the details of the model.

\subsection{Setup}

We adopt the same simplified set-up of T18 inspired by the results of PIC simulations for trans-relativistic magnetized shocks (the geometry is sketched in Fig. \ref{fig:cartoon}). We assume that the shock front is perpendicular to the jet axis (therefore the field parallel to the shock normal is the poloidal field and the perpendicular component is the toroidal one) and that it encompasses the entire jet cross section of the cylindrical jet. We assume that the downstream reference frame moves with Lorentz factor $\Gamma_{\rm d}$ with respect to the observer  and that the observer line of sight forms and angle $\theta_{\rm v,obs}=1/\Gamma_{\rm d}$ with respect to the jet axis. In this configuration the observer receives the photons emitted at an angle $\theta_{\rm v}=\pi/2$ in the downstream plasma frame. In this geometry frequencies and times are transformed from the downstream frame to the observer frame by using the Doppler relativistic factor $\delta=\Gamma_{\rm d}$. In the following, if not explicitly noted, all physical quantities characterizing the system (with the exception of $\Gamma_{\rm d}$) are expressed in the downstream reference frame.

The configuration of the self-generated magnetic field is modeled with a cell structure, in which each cell represents a coherence domain. Each domain has a cylindrical shape aligned with the jet axis and is characterized by radius $r$ and height $h=r$. In each cell we specify the total magnetic field as the sum of a constant parallel (poloidal) field $B_{\parallel}=B_z$ and the orthogonal field $\mathbf{B_{\perp}}$, with components $B_x$, $B_y$. As indicated by PIC simulations, the orthogonal component rapidly decays after the shock front. We model the decay with distance using a simple power-law prescription for the allowed maximum value $B_{\perp,\rm max}$. Specifically, we assume $B_{\perp,\rm max}(z)=B_{\perp, 0}(z/z_{\rm sh})^{-m}$, where $z_{\rm sh}$ indicates the position of the front\footnote{We use this analytical form to minimize the number of free parameters. A more general expression would include the decay length of the self-generated magnetic field, $z_{\rm decay}$: $B_{\perp,\rm max}(z)=B_{\perp, 0}[1+(z-z_{\rm sh})/z_{\rm decay}]^{-m}.$}. In each domain $B_{\perp}$ is randomly selected using a flat probability distribution in the interval 
$(0, B_{\perp, \rm max})$.
We then select the $x$ and $y$ components assuming a flat probability distribution for the angle $\alpha= \tan^{-1} (B_y/B_x)$.


Injected electrons follow a power law energy distribution $N(\gamma) \propto \gamma^{-p} $with slope $p=2$. Furthermore, we assume that the (possibly modulated) particle injection at the shock front is characterized by a timescale $t_{\rm dur}$. Specifically, our injection term will be:
\begin{equation}
\frac{dN(\gamma)}{d\gamma}=Ke^{-t/t_{\rm dur}}\gamma^{-2}, \;\;\; \gamma<\gamma_{\rm max,0}
\end{equation}
where $K$ is a constant.
Particles are advected downstream with constant comoving velocity $v_{\rm adv}\approx c/3$ (we neglect diffusive effects). Due to radiative losses, the energy of the electrons decreases with time (we ignore possible reacceleration due to turbulence or reconnection) and therefore with  distance from the shock. At a given distance from the shock, $z$, the evolved distribution will have the same slope, but the maximum energy will change with distance, $\gamma_{\rm max}(z)$. Since the total average magnetic field is the sum of the (constant) parallel and the (decreasing) perpendicular components, $B(z)^2=B_{\parallel}^2+B_{\perp}^2(z)$, the maximum Lorentz factor of the electrons at each distance is described by the differential equation $mc^2 d\gamma/dt =-(4/3) \sigma_T c U_B \gamma^2$ with the substitution $t \to z/v_{\rm adv}$ and $U_B=B(z)^2/8\pi$ (adiabatic and inverse Compton losses are neglected). 
We assume a jet radius $r=10^{15}$ cm and a downstream region extending from $z=10r=z_{\rm sh}=10^{16}$ cm to $z_{\rm max}=3\times z_{\rm sh}$. After $z_{\rm max}$ we assume that adiabatic losses totally quench the emission.

\begin{figure}
 \hspace*{-1.2truecm}
 \vspace*{-0.95truecm}
 \includegraphics[width=0.59\textwidth]{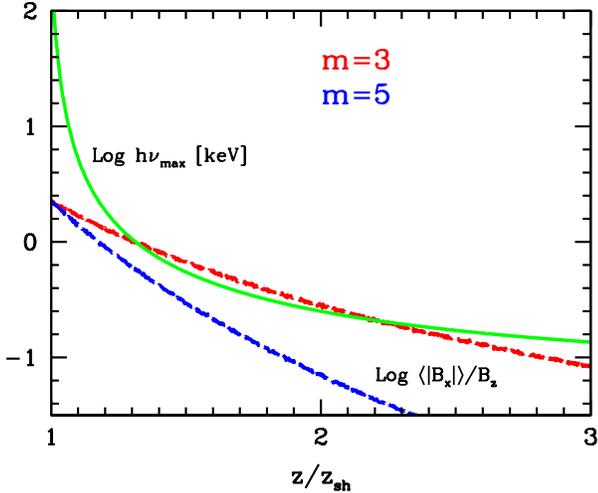}\\
 \vspace*{-4.1truecm}
 \caption{The solid curves report the logarithm of the ratio between $\langle |B_x| \rangle$, the cell-averaged $|B_x|$ (i.e. the perpendicular component of the magnetic field projected onto the plane of the sky), and the parallel magnetic field component as a function of the distance from the shock front for the physical setup used in the paper. The profiles correspond to $m=3$ (red) and $m=5$ (blue). The solid curve shows the maximum synchrotron photon energy (in the observer frame) as a function of the distance from the shock front. The curves correspond to the set of parameters given in the text. The radiation in the hard X-ray range ($h\nu_{\rm max} =10$ keV) is  produced in the region where $\langle |B_x| \rangle \gg B_{\parallel}$.}
 \label{fig:profiles}
\end{figure}

As in T18 we use as benchmark values $B_{\parallel}=0.015$ G and a ratio for the energy density of the perpendicular and parallel components at the shock $B_{\perp}^2/B_{\parallel}^2=50$. In this condition the initial total magnetic field is $B_0=0.12$ G\footnote{Note that, for simplicity we are assuming that the upstream field is exactly perpendicular to the shock normal, although only nearly perpendicular fields are required for efficient particle acceleration (e.g. Sironi et al. 2015).} This is the field surviving at some distance from the shock front, in regions where the self-generated field has already decayed.
The profiles of the ratio between $B_{x}(z)$ (averaged over the cells), which represents $B_{\perp}$ projected onto the plane of the sky, and $B_z=B_{\parallel}$, the parallel component of the magnetic field, and those of the maximum synchrotron frequency are shown in Fig. \ref{fig:profiles}. Note that for particles emitting at the highest energies (10 keV), the cooling length is of the order of $\simeq 0.1z_{\rm sh}=r/c$.

The geometrical treatment of the light-crossing time effects is implemented following the approach described in Chiaberge \& Ghisellini (1999). We refer the interested reader to the original paper for details.  
  
Knowing the magnetic field components for each domain $i$, we derive the frequency dependent Stokes parameters for synchrotron emission in the {\it observer} frame, $U_{\nu,i}$, $Q_{\nu,i}$ and $I_{\nu,i}$ (e.g. Lyutikov et al. 2005). Finally, the total observed degree of polarization, $\Pi_{\nu}$, and the electron vector position angle (EVPA), $\chi_{\nu}$, are derived from the total Stokes parameters $U_{\nu}=\sum U_{\nu,i}$,  $Q_{\nu}=\sum Q_{\nu,i}$ and $I_{\nu}=\sum I_{\nu,i}$ (where the cells to be summed are prescribed by the light-crossing time effects), by using the standard formulae $\Pi_{\nu} = \sqrt{Q_{\nu}^2+U_{\nu}^2}/I_{\nu}$ and
\begin{equation}
\cos 2\chi_{\nu}=\frac{Q_{\nu}}{\sqrt{Q_{\nu}^2+U_{\nu}^2}}, \;\; \sin 2\chi_{\nu}=\frac{U_{\nu}}{\sqrt{Q_{\nu}^2+U_{\nu}^2}}.
\end{equation}

\section{Results}

\begin{figure}
 \hspace*{-1.0truecm}
 \vspace*{-0.8truecm}
 \includegraphics[width=0.58\textwidth]{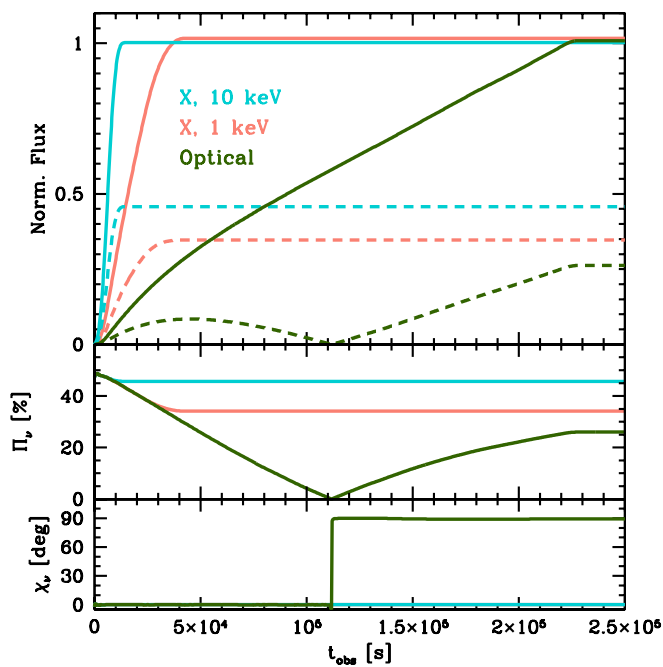}\\
 \vspace*{-1.5truecm}
 \caption{Upper panel: normalized light curves at 10 keV (light blue), 1 keV (orange) and in the optical band (green) assuming constant injection of particles starting at $t=0$ and $m=3$. The dashed line shows the polarized flux. Middle panel: degree of polarization in the three bands. Lower panel: EVPA in the three bands. All quantities are expressed in the observer frame.}
 \label{fig:pim3stat}
\end{figure}

In Fig. \ref{fig:pim3stat} we show an example with $m=3$ in which the injection is constant, i.e. $t_{\rm dur}\to \infty$. The upper panel shows the (normalized) lightcurves in three different bands, namely optical, soft X-rays (1 keV) and hard X-rays (10 keV). The dashed lines show the polarized flux, while the solid lines are for the total flux. The latter curves display a monotonic increase followed by a plateau, reached earlier for higher frequencies. The polarized flux for the X-ray band follows the same profile, while for the optical the situation is more complex. The dynamics can be understood considering in detail also the evolution of the degree and the angle of polarization (middle and lower panel, respectively).

The behavior of the X-ray emission can be simply understood. In fact, high-energy electrons radiating at these energies cool fast and their emission is therefore limited to a short layer after the shock (see Fig. \ref{fig:profiles}). The initial transient phase corresponds to the time required to fill the entire cooling length. After this phase the system reaches a stationary equilibrium between continuous injection and cooling. Since the emission region is limited to a thin layer, dominated by the orthogonal self-generated field (dashed line in Fig.\ref{fig:profiles}), the resulting radiation is highly polarized ($\Pi \approx$ 40\%) and the EVPA is close to 0 (corresponding to magnetic field lines mainly orthogonal to the jet axis). 

The optical band displays a more complex evolution. The cooling length of the electrons emitting at these frequencies corresponds to the entire emitting region $z_{\rm sh}-3z_{\rm sh}$ (see Fig. \ref{fig:profiles}). In the initial phase the electrons start to fill the regions close to the shock, implying a large degree of polarization, closely similar to that displayed by the X-ray band. While electrons are advected in regions where the orthogonal field is weaker and weaker, the total flux increases (the emitting volume steadily increases), while the polarized fraction diminishes. At a given point the integrated emission from the regions where  $\langle |B_x| \rangle>B_{z}$ (to simplify notation hereafter we write $\bar{B}_x$ instead of $\langle |B_x| \rangle$) exactly balances that from the more distant regions characterized by $\bar{B}_x<B_{z}$, determining a total degree of polarization $\Pi=0$.  After that point, the emission from the early injected electrons, advected by the flow in the regions dominated by the parallel field, starts to provide the largest contribution to the total emission. This time is marked by a sudden rotation of the EVPA of $\Delta \chi =90$ degrees and the increase of the  degree of polarization, that eventually becomes stationary at late times, when electrons reach a distance corresponding to $z_{\rm max}$ and fill the entire available volume. This final stationary state corresponds to the time-independent case discussed in T18.

In this example the degree of polarization in the X-ray band is basically constant and quite high, around 40\%, slightly smaller for 1 keV (since the corresponding electrons explore a region characterized by a smaller average orthogonal field with respect to those emitting at 10 keV). On the other hand, in the optical band the polarization is characterized by large variations. High polarization ($\Pi\gtrsim 20\%$) is in principle detectable only in the early phases.

\begin{figure*}
 \centering
 \hspace*{-1.3truecm}
 \includegraphics[width=1.05\textwidth]{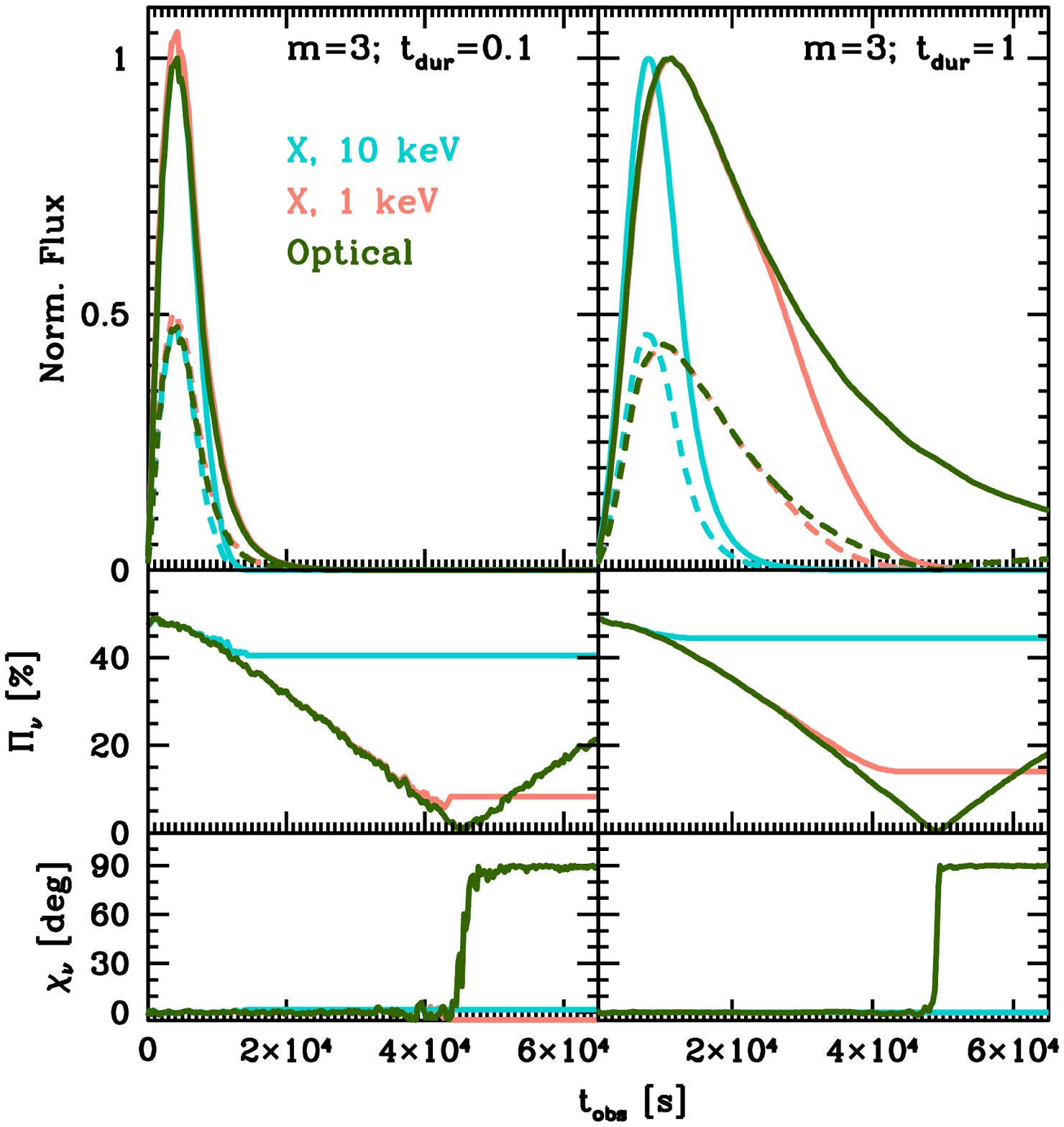}
 \vspace*{-0.8truecm}
 \caption{As in Fig. \ref{fig:pim3stat} for $m=3$ but for an injection timescale $t_{\rm dur}=0.1\times r/c$ (left) and $t_{\rm dur}=r/c$ (right).}
 \label{fig:pm3combo}
\end{figure*}

\begin{figure*}
 \centering
 \hspace*{-0.3truecm}
 \includegraphics[width=1.05\textwidth]{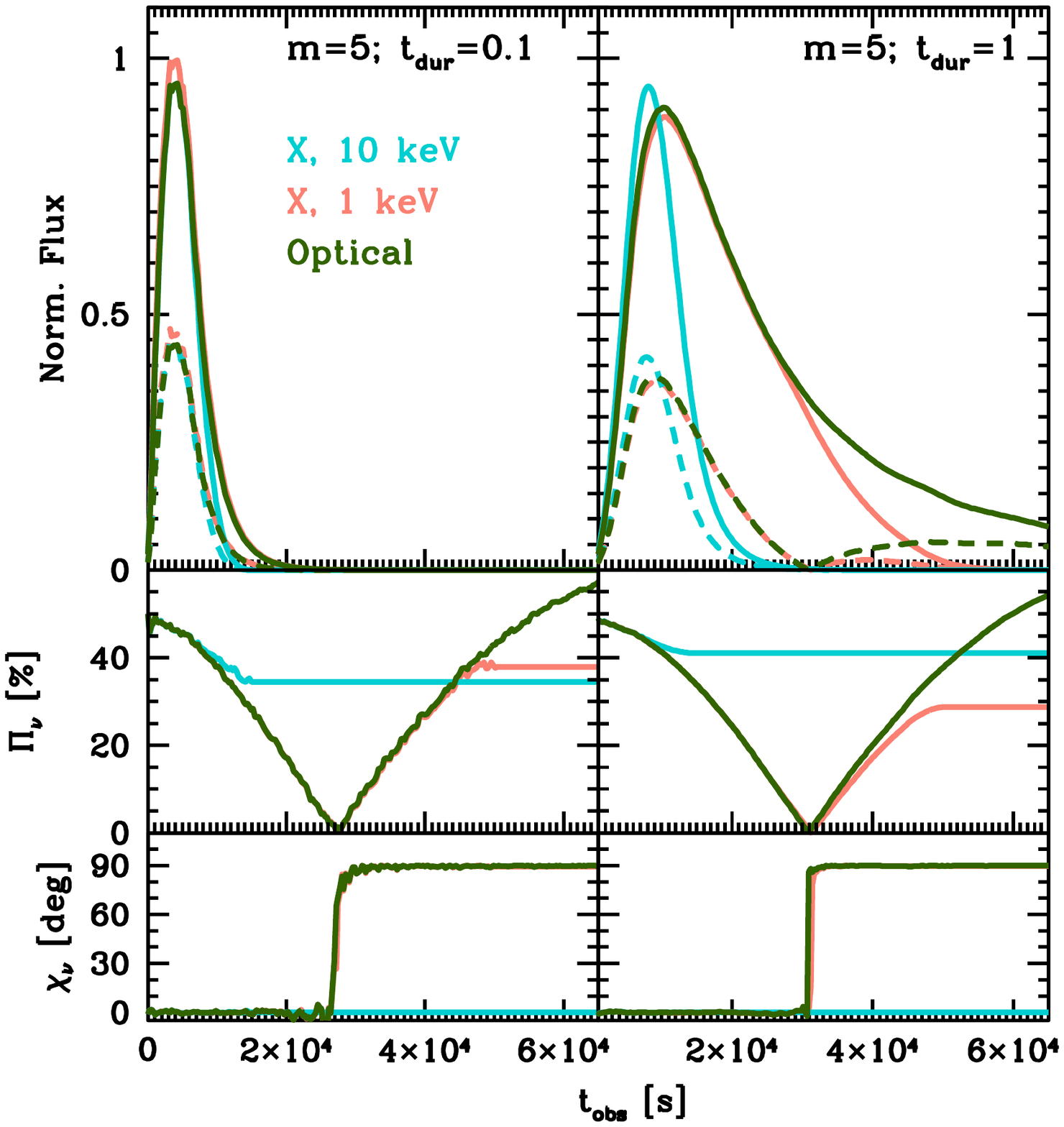}
 \vspace*{-1.5truecm}
 \caption{As in Fig. \ref{fig:pm3combo} but for $m=5$.}
 \label{fig:pm5combo}
\end{figure*}

In Fig.\ref{fig:pm3combo} we report two cases for $m=3$ with a finite injection timescale (expressed in units of $r/c$) of $t_{\rm dur}=0.1$ and 1. We report again three sets of curves corresponding to observed frequencies in the hard and soft X-ray band and in the optical band. 

Let us consider first the case with the shortest injection time, $t_{\rm dur}=0.1 r/c$ (left). In this case the shape of the flare in the three bands is quite similar (upper panel). In fact, since the injection is impulsive, the total observed duration of the flare is determined in this case by the light crossing time of the jet, so that both the rise and the decay timescale is $t_{\rm obs}\approx r/c\delta\simeq 3\times 10^3$ s. On the other hand, the properties of the polarization are rather different. For the case of 10 keV the degree of polarization starts around 50\% and after a small decrease stabilizes around 40\% (at times for which, however, the emitted flux is already low). For 1 keV and the optical  the situation is markedly different, since the degree of polarization displays a monotonic decay until a stationary state at $\Pi\sim 10\%$. The optical band follows closely the soft X-ray band but the decrease of the polarization fraction continues until it reaches zero, in correspondence to a sudden rotation of the polarization angle, which is then followed by an increase of $\Pi$. Again, as above, these late phases are of limited interest from an observational point of view, since they correspond to negligible levels of flux. However, it is interesting to understand the origin of this behavior, since, as we will see below, it can be relevant in other situations.

\begin{figure}
 \hspace*{-1.0truecm}
 \vspace*{-1.0truecm}
 \includegraphics[width=0.58\textwidth]{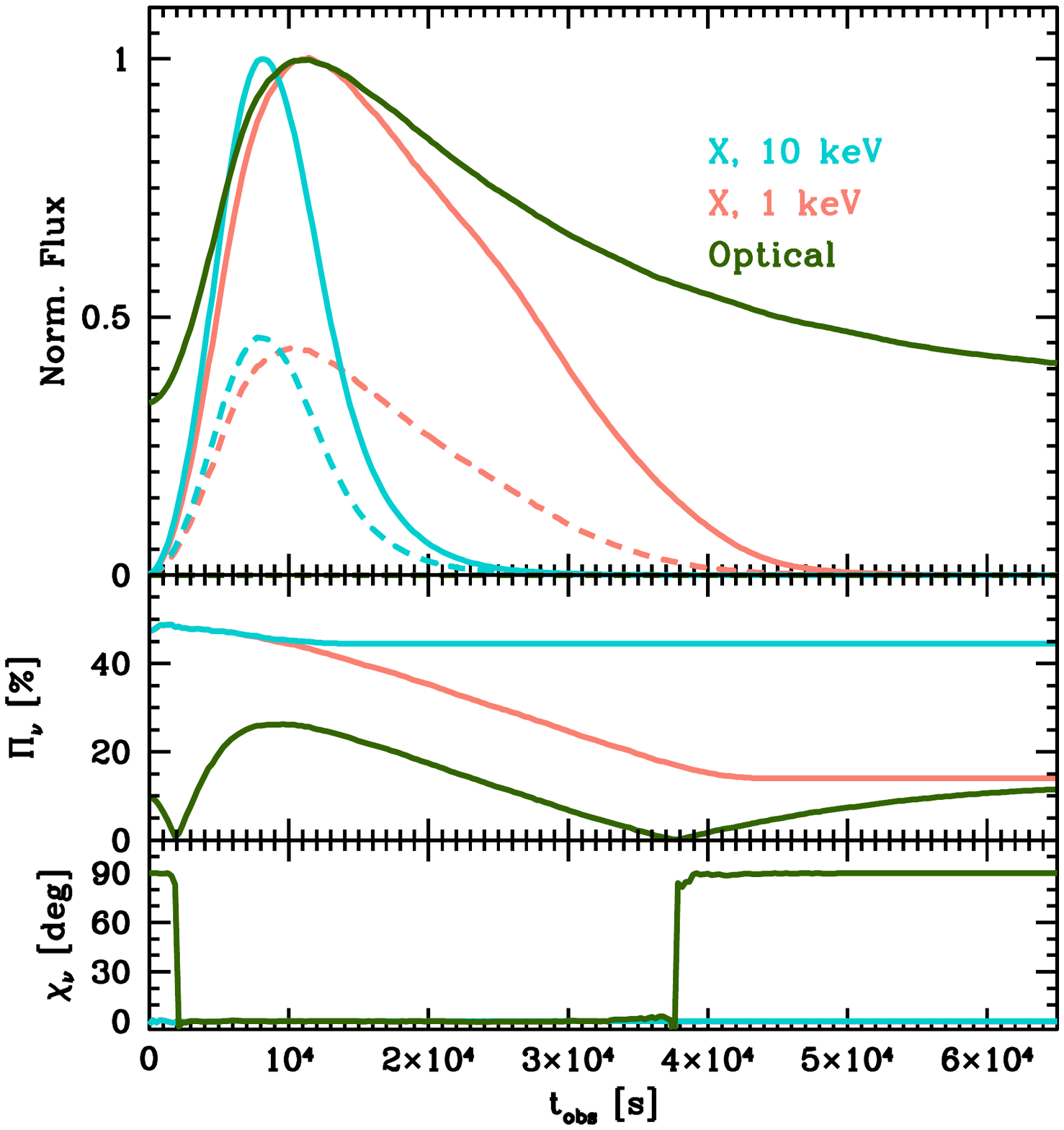}\\
 \vspace*{-1.5truecm}
 \caption{As for Fig.\ref{fig:pm3combo} (right) but including a diluting optical component. See text for details.}
 \label{fig:pim3dil}
\end{figure}

For the considered $t_{\rm dur}$, the emitting electrons fill a very thin ($\Delta z\simeq t_{\rm dur} v_{\rm adv}\simeq 3\times 10^{13}$ cm) layer which is carried by the flow. For the electrons at the highest energies the cooling length is short (e.g. less than $10^{15}$ cm for electrons emitting at an observed frequency of 10 keV) and therefore the emission from the layer switches off very close to the shock front. The polarization properties of the emerging radiation are therefore quite similar to those of the stationary injection case we discussed before. On the other hand, for electrons producing synchrotron photons at 1 keV and optical, the situation is different. Since the cooling length is appreciably larger than before, the layer filled by these electrons can travel for a sizeable distance from the shock front and therefore probe regions characterized by smaller values of the ratio $\bar{B}_x/B_{z}$, explaining the decrease of the polarized fraction. The electrons responsible for the emission at 1 keV cool before the distance where the projected perpendicular and the parallel magnetic field components are equal. On the other hand, the low-energy optical electrons survive well after this point. The transition from regions with $\bar{B}_x/B_{z}>1$ to those with $\bar{B}_x/B_{z}<1$ is marked by the rotation of the EVPA and the increase of $\Pi$ after the minimum at 0.

\begin{figure}
 \hspace*{-1.0truecm}
 \vspace*{-1.0truecm}
 \includegraphics[width=0.58\textwidth]{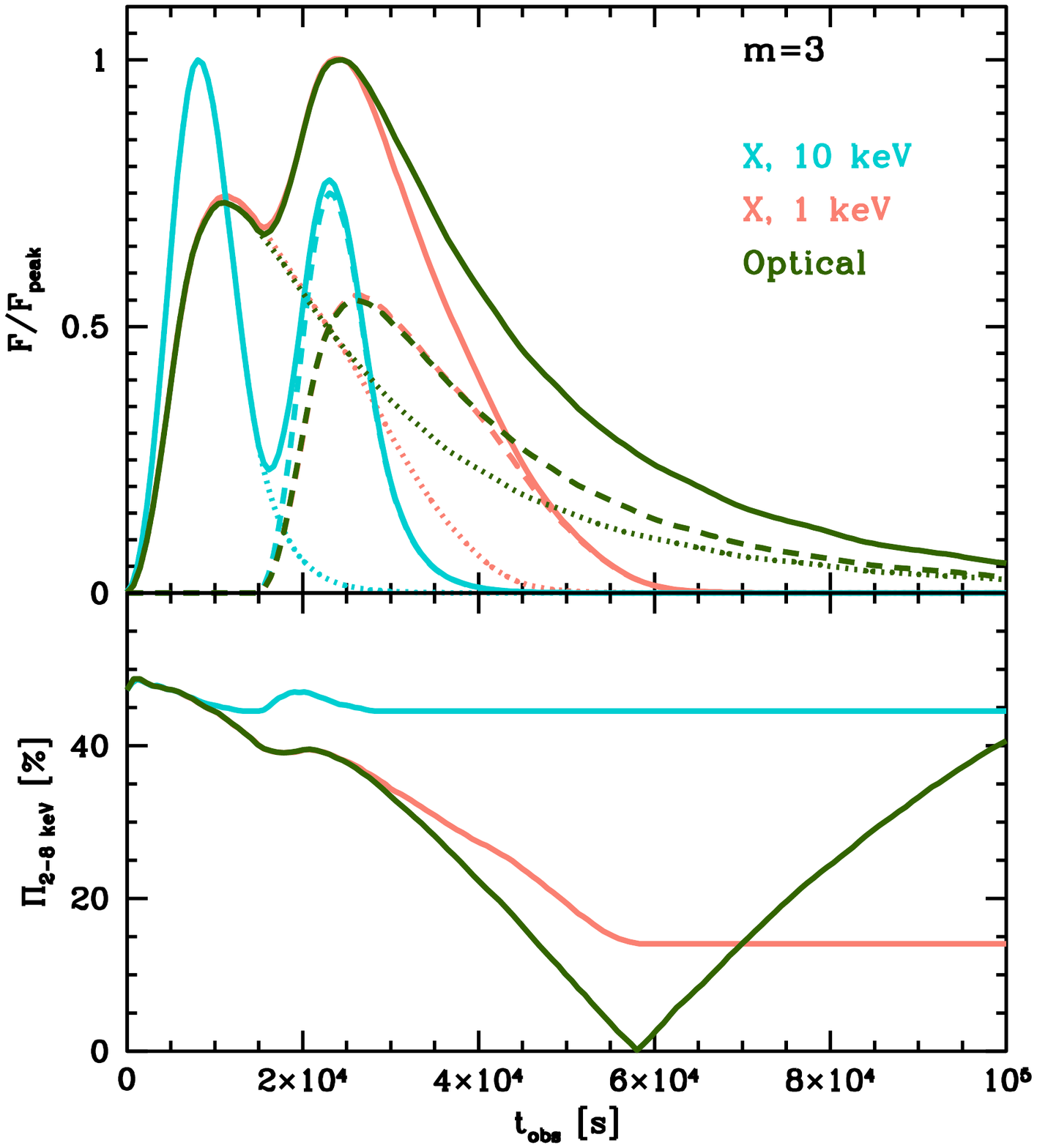}\\
 \vspace*{-1.5truecm}
 \caption{Normalized flux (upper panel) and degree of polarization (lower panel) for three different frequencies (10 keV, 1 keV, optical band) assuming two flares with duration $t_{\rm dur}=r/c$ separated by $\Delta t=1.5\times 10^5$ s (in the jet frame). The normalization of the second peak is fixed to 75\% of that of the first one.}
 \label{fig:pim3multi}
\end{figure}

The case in which the injection lasts for $t_{\rm dur}=r/c$ displays quite similar properties (right panel of Fig. \ref{fig:pm3combo}). However, in this case the decay of the flare lasts for a relatively longer time, so that the flux during the late phases is still appreciable, so the effects described above should be detectable.

In Fig. \ref{fig:pm5combo} we report two cases analogous to those in Fig. \ref{fig:pm3combo} but with $m=5$, thus characterized by a more rapid decay of the self-generated component of the magnetic field. We can observe the same qualitative behavior already discussed for the $m=3$ case. However, in this case the distance $z_{\rm eq}$ where $\bar{B}_x=B_{z}$ is closer to the shock than before. In this case, besides the low energy electrons radiating in the optical band, also the electrons producing the radiation at 1 keV reach $z_{\rm eq}$ before cooling and the polarization properties follow those of the optical emission, i.e. a rotation by 90 degrees of the EVPA and a degree of polarization that goes to zero and then increases. Note that in the case of the optical band, at the latest stages the electrons are embedded in a dominating parallel field, determining a degree of polarization even larger than that displayed by the X-rays. Again, these effects could be observable if the injection phase lasts for a time sufficient to guarantee an appreciable flux at late times (top panel).

In all previous examples we assumed that the observer detects only the emission produced in the downstream region of a single shock in the jet. However it is likely that, especially at the lowest frequencies (so, longest cooling times), the observed flux receives the contribution of a more extended portion of the jet (see e.g. Lindfors et al. 2016). As an example of a simple model for this situation we assume that, besides the optical emission produced by the electrons accelerated at the shock, there is a steady contribution characterized by a flux corresponding to 1/10 of the peak flux of the optical from the shock and with a polarization characterized by a low degree of $\Pi_{\rm dil}=10\%$ and an angle $\chi_{\rm dil}=90$ deg, corresponding to the ordered (i.e., not self-generated) poloidal field in the jet. The results, corresponding to $m=3$ and $t_{\rm dur}=r/c$, are shown in Fig.\ref{fig:pim3dil}. By construction, the X-ray lightcurves are the same as in Fig.\ref{fig:pm3combo} (right panel). The presence of the diluting component, on the other hand, has a substantial role in determining the polarization properties of the optical emission. Since the EVPA of the diluting component is orthogonal to that of the optical from the shock at early times, the degree of polarization shows an initial decrease, until the contribution of the shock emission exceeds that of the diluting component. At this point the angle changes by 90 degrees and the polarization degree starts to increase. At late times the diluting component becomes predominant again, determining the rotation of the angle by $\Delta \chi=90$ degrees and the increase of $\Pi$. A quite important difference with the case shown in Fig.\ref{fig:pm3combo} is that here the degree of polarization in the optical band hardly exceeds 20\%, in agreement with the observational evidence (e.g. Pavlidou et al. 2014, Covino et al. 2015).

In realistic situations, multiple flares can also occur with short time separation $\Delta t$. An example of the expected phenomenology in the simplest case of two injection phases is reported in Fig.\ref{fig:pim3multi}, showing (for the case $m=3$) the normalized flux (in the upper panel) and the degree of polarization (lower panel) calculated for two flares with duration $t_{\rm dur}=r/c$ separated by $\Delta t=1.5\times 10^4$ s (observer frame; this corresponds to 5 times $t_{\rm dur}/\delta$, i.e. the injection time in the observer frame) and a relative normalization between the second and the first flare of 0.75. The narrow peaks at 10 keV are well separated in the light curves, while the longer cooling times of electrons emitting at soft X-rays and in the optical band determines the blending of the two components at these energies. In all cases the degree of polarization shows a trend close to that of a single flare apart from a small bump in coincidence with the second flare. 

\subsection{Application to {\it IXPE}}

\begin{figure}
 \hspace*{-1.0truecm}
 \vspace*{-0.5truecm}
 \includegraphics[width=0.58\textwidth]{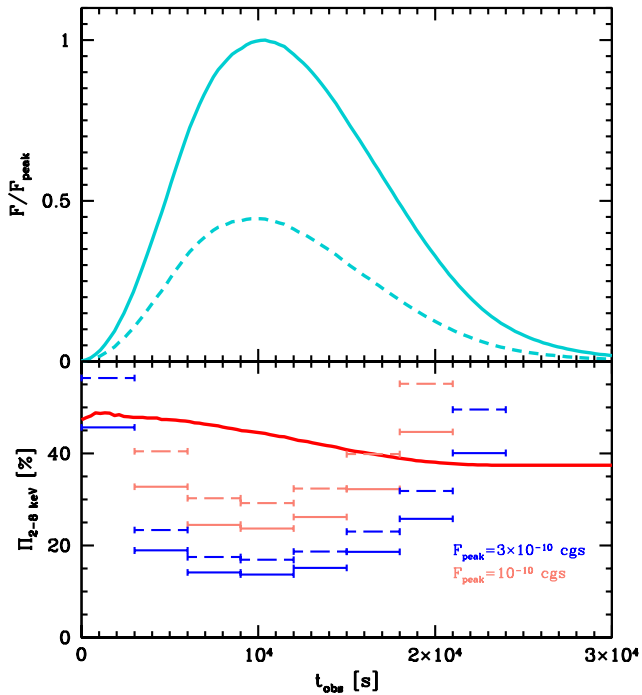}\\
 \vspace*{-1.5truecm}
 \caption{Lightcurve (upper panel) and degree of polarization (lower panel) in the {\it IXPE} band (2-8 keV) for the case $m=3$ and $t_{\rm dur}=r/c$. In the lower panel we show the minimum detectable polarization (MDP) at 99\% confidence for exposures of 1 ksec, for two peak fluxes and spectral photon index (1.5, solid; 3, dashed).}
 \label{fig:pim3ixpe}
\end{figure}

It is interesting to compare the predictions made by our scenario with the sensitivity foreseen for {\it IXPE}. An example is shown in Fig. \ref{fig:pim3ixpe} where we show the case with $m=3$ and $t_{\rm dur}=r/c$. In the upper panel we report the light curve in the {\it IXPE} band (2-8 keV) normalized to the peak flux and the polarized flux (dashed). In the lower panel we show the degree of polarization (averaged over the  {\it IXPE} band) as a function of time. For comparison we report the minimum detectable polarization (MDP) in time bins of 1 ksec for two different peak fluxes, $F_{\rm peak,2-8}=10^{-10}$ erg cm$^{-2}$ s$^{-1}$ (orange) and $F_{\rm peak,2-8}=3\times 10^{-10}$ erg cm$^{-2}$ s$^{-1}$ (blue)\footnote{We calculate the 99\% confidence MDP using the webPIMMS tool {\tt https://wwwastro.msfc.nasa.gov/ixpe/for\_scientists/pimms/} and then we scale it according to the assumed observation time.}. In both cases we show the MDP calculated assuming two different spectral photon index, $\Gamma_X=1.5$ (solid) and $\Gamma_X=3$ (dashed). It is clear that with the assumed flux the evolution of the degree of polarization could be tracked with great detail up to the end of the flare. More detailed simulations are beyond the scope of this paper.

We remark that the good sensitivity of {\it IXPE} will also make possible for sources with high flux to track the evolution of the polarization expected in case of multiple flares, as the case shown above (see Fig. \ref{fig:pim3multi}).

\section{Discussion}

We have developed a simple time-dependent scenario suitable to model the polarization signatures expected for the synchrotron radiation emitted by electrons accelerated at a trans-relativistic shock front in a blazar jet. The model is especially suitable to reproduce the phenomenology of highly peaked BL Lac objects, whose synchrotron emission extends up to the hard X-ray band. 

As already anticipated by using the stationary case presented in T18, a high degree of polarization is expected in the X-ray band. In fact, the X-ray radiation is produced by rapidly cooling electrons very close to the acceleration zone, where the magnetic field is dominated by the orthogonal self-generated field. On the other hand, the optical emission, produced by slowly cooling electrons, extends to regions far from the shock front, where we expect a weak magnetic field carried by the flow from upstream.

We fix the ratio between the self generated field and the field carried from upstream to a value consistent with that found in PIC simulations of trans-relativistic shocks. In this set-up the precise evolution depends mainly on two parameters, namely $m$, the index of the power law modeling the decay of the self-generated magnetic field in the downstream region, and $t_{\rm dur}$, the duration of the injection phase. We have shown results for slow and fast decay ($m=3$ and $m=5$) and impulsive ($t_{\rm dur}=0.1 r/c$) and prolonged ($t_{\rm dur}=r/c$) injection. In all cases the degree of polarization in the X-ray band settles at a relatively large value, while in the optical band the situation is more complex and the degree of polarization, as well as the polarization angle, shows a richer dynamics. 

In the optical band, our model predicts, on average, a large degree of polarization, in excess of $20\%$. However, it is well known that blazars, and HBLs in particular, show quite small polarization, at levels usually below $\Pi=10\%$ (e.g. Tommasi et al. 2001, Barres de Almeida et al. 2010, Pavlidou et al. 2014, Covino et al. 2015, Fraija et al. 2017). As suggested by multifrequency monitoring, at optical frequencies it is likely that the observed emission receives the contribution of an extended region of the jet, not limited to the active shock we are considering. This has been clearly demonstrated for the archetypal HBL Mkn 501 (Lindfors et al. 2016). In these conditions the observed polarization is no longer related to the geometry of a single region and a proper modeling would need a much complex set-up. This fact should be kept in mind when the polarization measured at low frequencies is extrapolated to predict the polarization in the X-ray band (e.g. Liodakis et al. 2019).

An important caveat to our model concerns the idealized set-up we have adopted in our study. In particular we assumed that the magnetic field upstream of the shock is completely parallel to the flow and that the shock normal is oriented along the flow direction. However, shocks in jets can easily develop with a non-negligible inclination with respect to the flow velocity -- typically if shocks derive from jet re-collimation (e.g. Komissarov 1994). Another important factor to take into account is turbulence in the flow, both upstream and downstream the shock (e.g. Baring et al. 2017). As discussed in Marscher (2014), the presence of turbulence would play an important role in shaping the polarization of the outcoming radiation. Moreover, if strong turbulence develops in the downstream flow, re-energization of low-energy particles can occur, modifying our results. Dedicated large scale kinetic shock simulations would be needed to assess the origin and strength of the turbulence, and its capabilities to reaccelerate particles far downstream from the shock.

The scenario we have considered, involving a trans-relativistic shock, can work only in the case of flows with low magnetization. In fact, as discussed in e.g., Sironi et al. (2015), high magnetizations hamper the efficient shock acceleration of relativistic particles\footnote{This challenges the results of Zhang et al. (2016) that argue that only shocks in a highly-magnetized flow can reproduce the observed phenomenology.}. Instead, at high magnetizations the most likely mechanism at work would be magnetic reconnection, triggered, e.g., by the onset of kink instability (e.g. Begelman 1998, Giannios \& Spruit 2006, Barniol Duran et al. 2017, Davelaar et al. 2020). In this context the simulations reported by Zhang et al. (2017) show that in the early phases of the development of the instability, when few regions are active, the degree of polarization can be relatively large $\Pi\sim 30-40\%$. On the other hand, while the instability develops and involves a larger portion of the jet, the polarization fraction tends to decrease. The polarization angle, on the other hand, displays a quite erratic behavior. An analysis of the polarization expected from a kink unstable jet, particularly focused on the differences between low frequencies (i.e. optical band) and the X-ray band is presented in Bodo, Tavecchio \& Sironi (2020).

The large degree of polarization expected for HBL in the X-ray band, coupled with the large flux displayed by the brightest sources of this class (exceeding $10^{-10}$ erg cm$^{-2}$ s$^{-1}$) would allow the upcoming satellite {\it IXPE} to track in detail the evolution of the polarization parameters ($\Pi$, $\chi$), during a fair fraction of a flare duration. Complemented with polarimetric observations in the optical band, such data would allow us an unprecedented view on the structure of trans-relativistic shocks and the ongoing particle acceleration in relativistic jets.

\section*{Acknowledgments}
We thank the referee, Alan Marscher, for useful comments. FT acknowledges contribution from the grant INAF Main Stream project "High-energy extragalactic astrophysics: toward the Cherenkov Telescope Array". LS acknowledges support from the Sloan Fellowship, the Cottrell Fellowship, NSF AST-1716567, NASA ATP NNX17AG21G and NSF PHY-1903412.

\section*{Data availability}

Data available on request.

\appendix

\section{Numerical implementation}
As already discussed in T18, the computation of polarization degree, especially when time dependence is considered, is a task that requires an adequate amount of computational power that in this case is of the order of few tens of thousands core hours. To perform our simulations, we developed an ad-hoc numerical code in Python, following the same approach described in T18 (see \cite{landonia, landonib} for further details). 

\begin{figure*}
 \hspace*{-1.0truecm}
 \includegraphics[width=0.85\textwidth]{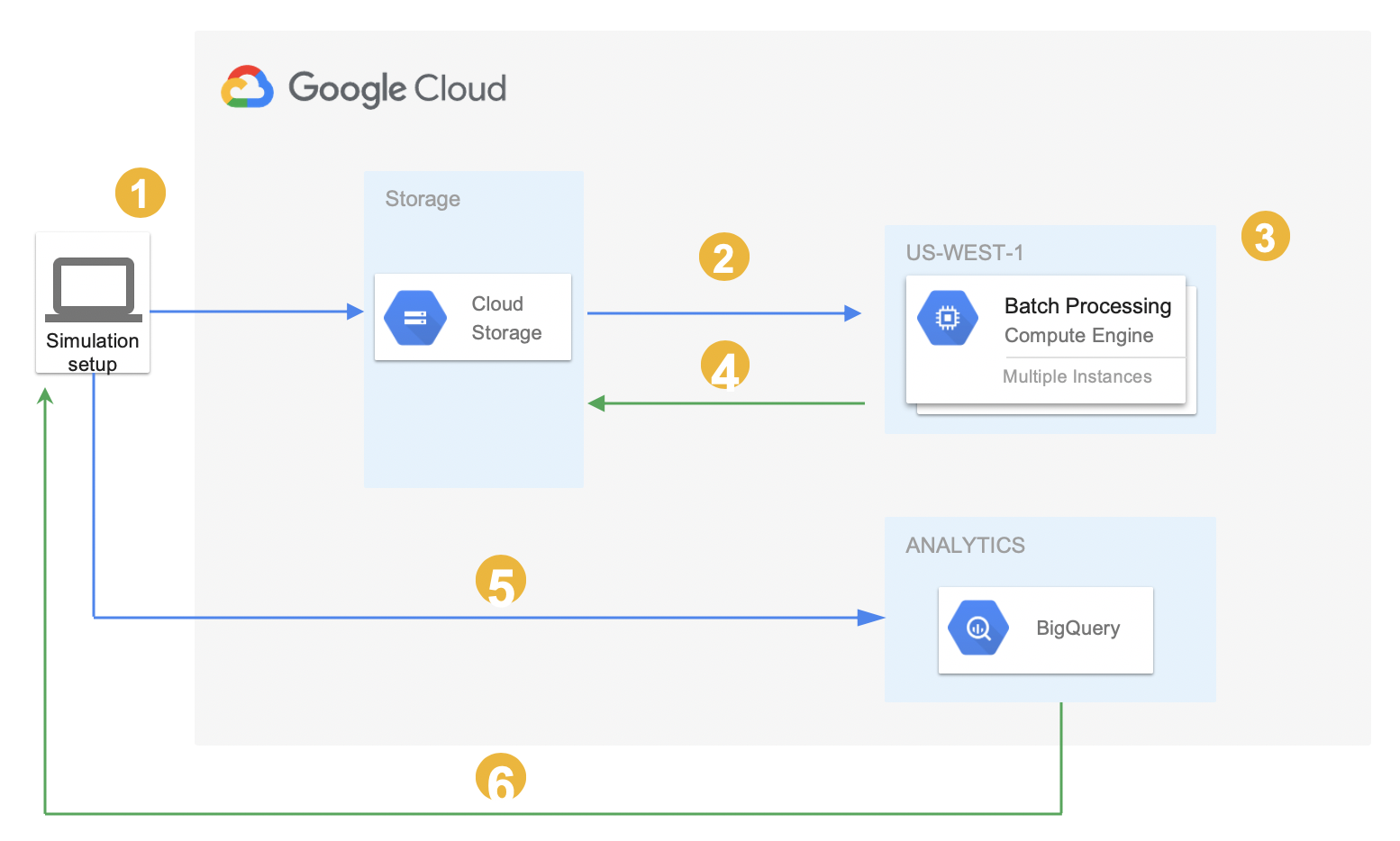}\\
 \caption{Cloud-based architecture adopted in the context of the implementation of numerical codes used in our model. We exploited the main services offered for computational power (Batch processing) and analytics (BigQuery) to speed up the computation and explore the parameter space. The computation starts at (1) where the input file is upload on the Cloud Storage. Then in (2) the file is splitted and the computation is distributed across many nodes (3). The intermediate results are stored back into the Cloud Storage (4) and by using BigQuery (5-6) the final Stokes parameters are computed and sent back for analysis. }
 \label{fig:cloud}
\end{figure*}

Briefly, we dived the program in two parts. The first one is in charge to produce the physical parameters of each cell (in space and time) while the second computes, for each frequency and for each cell, the associated Stokes parameters. 
In order to speed up the computation, we adopted the Cloud based architecture reported in Figure \ref{fig:cloud}. Since each cell is independent from each others, the whole input is divided in many different portions and computed by different nodes on the cloud that, at the end of their computation, store a text file that contains the relevant Stokes parameter of each cell on the storage.
The output files are then combined toghter and stored into a large no-SQL database (various hundreds GB
per each model) that is analysed using the cloud service BigQuery that implementes a Map-Reduce task to obtain the integrated Stokes parameters for
each frequency.

\end{document}